\documentclass[twocolumn,showpacs,preprintnumbers,amsmath,amssymb]{revtex4}
\usepackage{epsfig,amssymb,euscript} 
 \def\be{\begin{equation}}
 \def\ee{\end{equation}}
 \def\bea{\begin{eqnarray}}
 \def\eea{\end{eqnarray}}

\newcommand{\reef}[1]{(\ref{#1})}

\newcommand{\eg}{{\it e.g.,}\ }
\newcommand{\ie}{{\it i.e.,}\ }
\newcommand{\comment}[1]{{\bf [[#1]]}}

\def\nc{N} 

\def\cM{{{\cal M}_5}}
\newcommand{\ccct}{{C^3 \mathcal T}}

\begin{document}
\preprint{DAMTP-2008-48}
\title{
Quantum corrections to $\eta/s$
}
\author{Robert C. Myers$^{1,2}$}
\author{Miguel F. Paulos$^3$}

\author{Aninda Sinha$^1$}

\affiliation{$^1$ Perimeter Institute for Theoretical Physics, Waterloo, Ontario N2L 2Y5, Canada.
\\$^2$ Department of Physics and Astronomy, University of Waterloo, Waterloo, Ontario N2L 3G1, Canada.
\\$^3$Department of Applied Mathematics and Theoretical Physics, Cambridge CB3 0WA, U.K.}
\date{\today}
\begin{abstract}
We consider corrections to the ratio of the shear viscosity to the
entropy density in strongly coupled nonabelian plasmas using the
AdS/CFT correspondence. In particular, higher derivative terms with
the five-form RR flux, which have been ignored in all previous
calculations, are included. This provides the first reliable
calculation of the leading order correction in the inverse 't Hooft
coupling to the celebrated result $\eta/s=1/4\pi$. The leading
correction in inverse powers of the number of colours is computed.
Our results hold very generally for quiver gauge theories with an
internal manifold $L_{pqr}$ in the holographic dual. Our analysis
implies that the thermal properties of these theories will not be
affected by the five-form flux terms at this order.
\end{abstract}
\pacs{11.25.Tq, 12.38.Mh}
\maketitle

The gauge/gravity correspondence presents a powerful new framework
with which to study strongly coupled gauge theories \cite{adscft}.
One of the most striking new insights is that the ratio of the shear
viscosity $\eta$ to the entropy density $s$ is universal with
$\eta/s=1/4\pi$, for any gauge theories with an Einstein gravity
dual in the limit of a large number of colours and large 't Hooft
coupling, \ie $\nc,\lambda\rightarrow\infty$ \cite{universe}. In
fact, this result has been conjectured to be a universal lower bound
(the KSS bound) in nature \cite{KSS}. While this bound is far below
the ratio found for ordinary materials, a great deal of excitement
has been generated by experimental results from the Relativistic Heavy
Ion Collider suggesting that, just above the deconfinement phase
transition, QCD is very close to saturating this bound \cite{rick}.
Certainly to make better contact with QCD, it is important to
understand $1/\lambda$ and $1/\nc$ corrections which arise from
higher derivative corrections to classical Einstein gravity in the
holographic framework. For four-dimensional ${\cal N}=4$
super-Yang-Mills (SYM), the leading $1/\lambda$ correction has been
calculated as \cite{alex1}
 \be\label{hd}
\frac{\eta}{s}=\frac 1{4\pi}\,\left(1+{15\zeta(3)\over
\lambda^{3/2}} \right)\,.
 \ee
It has often been noted that with this correction added, the SYM
theory no longer saturates the KSS bound but given that the
correction is positive, the bound is still respected. However, as we
explain, the calculation producing \reef{hd} is incomplete and so
neither the precise coefficient nor even the sign of the correction
could be stated with any certainty.

In the above calculations, in order to introduce a finite
temperature, one considers the background of the AdS-Schwarzschild
black hole, $AdS_{BH}\times \cM$. In 10d, the fields involved are the metric and the
five-form:
 \bea
ds^2 &=& \frac{(\pi T L)^2}u (-f
dt^2+d \mathbf {x}^2)+\frac{L^2 du^2}{4u^2 f}+L^2d S_{\cM}^2
\nonumber \\
F_5 &=& -\frac{4}{L}(1+\star) \mbox{vol}_{\cM},\qquad \mathbf
{x}=(x,y,z) \,.\label{AdSBH}
 \eea
Here we have $f=1-u^2$ and $T$ is the the temperature of the black
hole. $L$ is the AdS radius of curvature with
$L^4=\lambda\,\alpha'^2$ and $\mbox{vol}_{\cM}$ is the volume-form
on the compact five-manifold $\cM$. Of course, the geometry dual to
${\cal N}=4$ SYM has $\cM=S^5$ but our discussion in the following
extends to more general internal manifolds.

The coupling constant dependence of thermodynamic quantities can be
obtained by considering higher derivative $\alpha'$ corrections to
the supergravity action and their effect on the geometry
(\ref{AdSBH}). The first corrections in the IIB theory arise at
$\alpha'^3$ relative to the supergravity action, which leads to
thermodynamic corrections suppressed by $1/\lambda^{3/2}$. The best
understood correction is the $R^4$ term \cite{Grisaru:1986px}, 
which by a field redefinition can be written in terms of the Weyl
tensor $C_{abcd}$
 \bea
S^{(3)}_{R^4}&=& {\gamma \over 16 \pi G}\int d^{10}x~ \sqrt{-g} ~
e^{-\frac 32 \tilde \phi}\, W_{C^4} \label{R4} \\
W_{C^4}&=& C_{a b c d} C^{a b}_{\ \ e f} C^{c \ d}_{\ g \ h} C^{e g
f h}-\frac 12 C_{a b c d} C^{a b}_{\ \ e f} C^{c e}_{\ \ g h} C^{d g
f h} \nonumber
 \eea
where $ \gamma=\frac 18 \zeta(3) \alpha'^3$. We have written
(\ref{R4}) in the Einstein frame and in the limit of small string
coupling. We have also introduced notation for the dilaton:
$e^{\phi}=g_s e^{\tilde \phi}$ where $g_s$ is the string coupling
and $\tilde \phi$ is the deviation from the asymptotic value. The
corrections to the thermodynamics of ${\cal N}=4$ SYM arising from
this term were first studied in \cite{Gubser:1998nz,Paw}. However,
this $C^4$ term is only one of a large collection of terms at this
$\alpha'^3$ order. In particular, there are terms involving the
five-form RR flux which cannot be ignored if one wishes to make
rigorous statements on the SYM thermodynamics. As such, all existing
results in the literature are incomplete.


A few years ago, Green and Stahn \cite{Green:2003an} proposed that
the IIB higher derivative terms involving the graviton and the
five-form flux can be completely specified. The fields in IIB
supergravity can be written compactly as an analytic superfield
\cite{Howe:1983sra} whose first component is $\tau=a+i e^{-\phi}$.
Higher derivative terms at $O(\alpha'^3)$ can be written as an
integral of functions of this superfield over half the superspace.
In particular, \cite{Green:2003an} proposed a supersymmetric
completion of $C^4$ in the case where only the five-form and metric
are present. This proposal was shown \cite{Rajaraman:2005up} to
evade potential problems in defining a supersymmetric measure
\cite{deHaro:2002vk}. Using this proposal, it was found
\cite{Green:2003an} that the D3-brane solution in supergravity does
not get renormalized by the higher derivative terms. This is a
useful check since it is expected that the near-horizon limit
AdS$_5\times$ S$^5$ is a solution to all orders in $\alpha'$
\cite{allorder}. In contrast, the potential pitfalls faced with incomplete
calculations without the five-form flux terms were demonstrated in \cite{sinkowrong}, where an illustrative analysis with the $C^4$ term alone produced an incorrect modification of the D3-brane solution.
More concretely, the correction determined by \cite{Green:2003an}
is
\begin{widetext}
 \be
S^{(3)}_{{\mathcal R}^4}= \frac {\alpha'^3g_s^{3/2}}{32 \pi G} \int
d^{10} x \int d^{16} \theta ~ \sqrt{-g}f^{(0,0)}(\tau,\bar \tau)
[(\theta \Gamma^{mnp}\theta) (\theta \Gamma^{qrs}\theta)\mathcal
R_{mnpqrs}]^4\ +\ c.c.\label{SUSYR4}
 \ee
\end{widetext}
where $f^{(0,0)}(\tau,\bar\tau)$ is a modular form \cite{greengutperle}. The latter
captures the behaviour for all values of the string coupling but
remarkably contains only string tree-level and one-loop
contributions, as well as an infinite series of instanton
corrections. The six-index tensor $\mathcal R$ is specified by \cite{deHaro:2002vk, Green:2003an}
 \be
\mathcal R_{mnpqrs}= \frac 18 g_{ps} C_{mnqr}+\frac i{48} D_m
F^+_{npqrs} + \frac 1{384} F^+_{mnptu}{F^+}_{\!\!\!\!\!\!qrs}^{\ tu}
 \label{DefR}\,,
 \ee
where $F^+=\displaystyle {1\over 2}(1+*)F_5$. Note that only having
this self-dual combination 
is essential to the $\alpha'$-corrected $F_5$ equation being cast
as a self-duality constraint \cite{pw}:
 \be\label{Feom}
\displaystyle F_5-120\gamma {\delta W_{{\mathcal R}^4}\over \delta F_5}=*
\left(F_5-120\gamma {\delta W_{{\mathcal R}^4}\over\delta F_5}\right)\,.
 \ee
Building on methods of \cite{Green:2005qr}, the specific tensor
structure of the terms contained in (\ref{SUSYR4}) was computed in
\cite{Paulos:2008tn}. The result is a set of 20 independent terms
%
which can be schematically written as
 \be \label{sch}
C^4+C^3 \mathcal T+C^2 \mathcal T^2 +C \mathcal T^3+\mathcal T^4\ +\
c.c.\,,
 \ee
where the tensor $\mathcal T$ is defined as
 \bea
\mathcal T_{a b c d e f}&\equiv& i \nabla_{a} F^+_{b c d e f}
 \label{DefT}\\
&&+\frac 1{16} (F^+_{a b c m n}{F^+}_{\!\!\!\!\!\!d e f}^{\  m n}-3
F^+_{a b f m n}{F^+}_{\!\!\!\!\!d e c}^{\ m n}) \,.\nonumber
 \eea
Implicit above are antisymmetry in $[a,b,c]$ and $[d,e,f]$, as well
as symmetry in the exchange of the triples, which arises from the
gamma-matrix structure in \reef{SUSYR4}. The form of \reef{SUSYR4}
also imposes a projection onto the irreducible $\bf{1050}^+$
representation of $\nabla F_5^+$ and $F_5^{+\,2}$. The particular
combination of $F_5^{+\,2}$ terms in \reef{DefT} affects this
projection, while for $F_5$ satisfying the leading supergravity
equation of motion $F_5=*F_5$, this projection reduces to the
identity on $\nabla F_5^+$.
To make \reef{sch} more explicit, the $C^4$ term 
reproduces \reef{R4} and the $C^3 \mathcal T$ term is written as
\cite{Paulos:2008tn}
 \be\label{corr1}
W_{C^3 \mathcal T} ={3\over 2}C_{a b c d}\, C^{a}_{\ e f g}\,
 C^{b f}_{\ \ h i}\, \mathcal T^{c d e g h i}\ +\ c.c.\,.
 \ee

We will now proceed to analyse the effect of these new terms on
$\eta/s$. Of course, their effect in calculating the shear viscosity
$\eta$ and the entropy density $s$ must be considered separately.
The latter requires determining the effect of the higher order terms
on the event horizon in the background solution
\cite{Gubser:1998nz,wald}. $\eta$ may be calculated by perturbing
the black hole background by a metric deformation $h_{\mu\nu}$ and
either applying Kubo formulae to correlators of the dual stress
tensor \cite{kubo} or examining the hydrodynamic form of the
resulting stress tensor \cite{hydro}. Again, we want to consider how
the $\alpha'^3$ terms modify this `deformed' solution. Hence in the
following, we focus on how the higher order terms effect the
equations of motion. Our conclusion is that the relevant backgrounds
receive no $\alpha'^3$ corrections from five-form terms!

We begin with a general supergravity solution of the form $A_5\times
\cM$. Our discussion will be general and we only assume that $A_5$
and $\cM$ are Einstein manifolds with equal and opposite curvatures. However, in our application below, we
intend that $A_5$ is an asymptotically AdS black hole (with or
without the deformation). Beyond the usual choice of $S^5$ for
$\cM$, our general discussion includes the Einstein-Sasaki manifolds
$L_{pqr}$ \cite{lpqr}, of which $Y^{p,q}$ and $T^{1,1}$ are special
cases. In the following, we denote the $A_5$ and $\cM$ coordinates
as $a_i, m_i$, respectively.
%
For consistency of the expansion in $\alpha'^3$, we need only use
this leading order supergravity solution in evaluating the
contributions of the higher derivative terms.

The key feature of the above background is that the five-form
solution remains as in \reef{AdSBH}, \ie it is the sum of the volume
forms on $A_5$ and $\cM$. This five-form solution immediately
implies that the tensor $\mathcal T$ vanishes. The $\nabla F^+_5$
piece vanishes because the volume forms are covariantly constant.
The particular combination of $F_5^{+2}$ terms appearing in
\reef{DefT} also vanishes in this case.
Since $\mathcal T=0$, the only term other than $C^4$ that could
affect the equations of motion is the $C^3 \mathcal T$ term, given
in \reef{corr1}.

To show that this term cannot affect the equations of motion, we
first note that the structure of \reef{corr1} yields the form
$P(C^3) P(F_5^{+2})$ where $P$ is the $\bf{1050}^+$ projection
operator, whose action on $F_5^{+2}$ was described above. On the
$C^3$ factor, the projection operator yields
 \be
P(C^3)_{a b c}^{\ \ \ d e f}=\frac 12 \left((C^3)_{a b c}^{\ \ \ d e
f}-3 (C^3)_{[a b \ \ \ \ c]}^{\ \ [f d
e]}\right)\,.\label{Projection}
 \ee
Now for our purposes, it suffices to consider the tensor $\delta
{W_{\ccct}}/\delta F_5=P(C^3)F_5$ evaluated on the background.
Firstly, since $A_5$ and $\cM$ are Einstein, the only
non-zero components of the Weyl tensor are $C_{a_1 a_2 a_3 a_4}$ and
$C_{m_1 m_2 m_3 m_4}$ \cite{comm1}.
Now in
$\delta {W_{F_5}}/\delta F_5$, $P(C^3)$ contracts into either
$F_5^{a_1 a_2 a_3 a_4 a_5}$ or $F_5^{m_1 m_2 m_3 m_4 m_5}$ and
since uncontracted indices are antisymmetric, the only relevant components
are $P(C^3)_{a_1 a_2 a_3 a_1 a_2 a_3}$ and $P(C^3)_{m_1 m_2 m_3 m_1
m_2 m_3}$.
%
However, using \reef{Projection}, a simple calculation explicitly
shows that in fact for an arbitrary tensor $M_{[abc][def]}$, $P(M)$
has no such components. Therefore $\delta {W_{\ccct}}/\delta F_5$ is
always vanishing for the relevant backgrounds. 
Given that $P(C^3) F_5=0=\mathcal T$, it also follows $\delta
{W_{\ccct}}/\delta g_{ab}$ vanishes. Hence we may conclude that
only $C^4$ alters the geometry at order $\alpha'^3$.

With this result, we may conclude that the $F_5$ terms in
\reef{SUSYR4} do not affect the ratio $\eta/s$. First of all, to
consider $\alpha'^3$ corrections to the asymptotically AdS black
hole in \reef{AdSBH}, the above proof indicates that these arise
only from the the $C^4$ term \reef{R4}. Hence we have rigorously
established that the entropy density for the SYM theory takes the form in
\cite{Gubser:1998nz,wald}.
Note the observation that only the $C^4$ terms correct the static
black hole geometry was made in \cite{Paulos:2008tn} but our
discussion extends this result to more general $\cM$.

Corrections to $\eta$ can be considered with $A_5$ being
asymptotically AdS black hole with an appropriate deformation
$h_{\mu\nu}$. Here again, our proof indicates that only the $C^4$
term is relevant and so we have rigorously established the finite
coupling corrections appearing in \reef{hd}. In fact, our results
here have wider significance for general holographic calculations of
such transport coefficients. As recently explored \cite{hydro}, the
hydrodynamics of the conformal plasmas are governed by a variety of
higher order coefficients as well as $\eta$. Our present discussion
indicates that it is sufficient to consider the $C^4$ term to
calculate the $\alpha'^3$ corrections to all of these coefficients
\eg the relaxation time \cite{relax}. We emphasise that the last
statement includes the coefficients for terms nonlinear in the local
four-velocity, as the above discussion applies to the deformed black
hole fully nonlinear in $h_{\mu\nu}$.

One might also consider the universality of the corrections to
transport coefficients proposed in \cite{universal}. There it was
found that the full spectrum of quasi-normal modes (and hence
implicitly the linear transport coefficients, such as $\eta$)
matches for $\cM=S^5$ and $T^{1,1}$. While it is beyond the scope of
the present article to determine whether or not this proposal is
correct, the present discussion indicates at the $F_5$ terms at
order $\alpha'^3$ are irrelevant to determining the spectrum.

Implicitly the above conclusions rely on these results not being
modified by new fields which are trivial in the original background.
In principle, other terms in the $\alpha'^3$ action beyond those
captured in \reef{SUSYR4} might source other type IIB fields. For
definiteness, consider the RR axion $a$ which vanishes at lowest
order. There might still be $\alpha'^3$ terms which are linear in
$a$, \eg $ C^2 \nabla F^+_5 \nabla^2 a$. The corrected solution
would then also include an axion of order $\alpha'^3$. However, in
Einstein's equations, $a$ will only appear quadratically or in terms
with an $\alpha'^3$ factor. Hence, its effects in, \eg the
quasi-normal spectrum will only be felt at order $\alpha'^6$ and
therefore can be neglected here. The same reasoning can be made for
all other fields. In particular, it applies to the warp factor which
is induced at order $\alpha'^3$ \cite{alex1}. This case is slightly
different in that this field is a new correction to the metric and
RR five-form. However, the same reasoning can be applied in the
effective five-dimensional theory on $A_5$, in which, to leading
order, the warp factor appears as a massive scalar field with
standard quadratic action.

Holographic analyses are typically applied with both
$\nc,\lambda\rightarrow\infty$. So it is of interest to consider
corrections for both finite $\nc$ and finite $\lambda$.
Higher-derivative corrections in string theory at perturbative level
in string coupling $g_s$ are weighted by $\alpha'^n g_s^{2m}$ with
$g_s=\lambda/4\pi N_c$ and so one expects that $\eta/s$ can be
written as
 \be\label{ser}
\frac{\eta}{s}={1\over 4\pi}\left(1+\sum_{\substack{n=3\\m=0}}
c_{nm} \lambda^{-n/2}\left[\left({\lambda\over 4\pi
N_c}\right)^{2m}+f_{NP}\right]\right)\,,
 \ee
where the non-perturbative contributions $f_{NP}$ come from
D-instanton effects. Remarkably the modular form
$f^{(0,0)}(\tau,\bar\tau)$ only contains two perturbative terms
\cite{greengutperle}:
 \be \label{sloop}
f^{(0,0)}_P= {\zeta(3)\over 8} e^{-3\phi/2} \left(1+{\pi^2\over 3
\zeta(3)}e^{2\phi}\right)\,.
 \ee
Since the dilaton only varies at $O(\alpha'^3)$, these contributions
extend \reef{hd} to
 \be
\frac{\eta}{s}=\frac 1{4\pi}\,\left(1+{15 \zeta(3)\over
\lambda^{3/2}}+ \frac{5}{16}{\lambda^{1/2}\over \nc^2}+\tilde
f_{NP}\right)\,. \label{hd2}
 \ee
Examining the nonperturbative contribution in detail yields $\tilde
f_{NP}\approx 15/(2 \pi^{1/2} \nc^{3/2}) e^{-{8\pi^2 \nc\over
\lambda}}$ for small $g_s$ \cite{greengutperle}. Hence these
contributions are subdominant in the full perturbative expansion
\reef{ser}. However, the second correction above is enhanced by a
factor of $\lambda^{1/2}$ and so this is the leading order
correction in inverse powers of $\nc$ for ${\cal N}=4$ SYM. The same
enhancement appears in the leading $1/\nc^2$
correction to the entropy density \cite{Gubser:1998nz}:
 \be
s= \displaystyle{\pi^2 \nc^2 T^3\over 2}\left(1+{15\over 8}
{\zeta(3)\over \lambda^{3/2}}+{5\over 128} {\lambda^{1/2}\over
\nc^2}  
\right)\,. \label{sent}
 \ee

One might be tempted to consider how these first two corrections
might modify $\eta/s$ in QCD's strongly coupled quark-gluon plasma.
Substituting  $\lambda=6\pi$ (\ie $\alpha_s= 0.5$) and $N_c=3$, they
increase $\eta/s$ from $1/4\pi\approx0.08$ to $0.11$. The first and
second terms produce roughly 22\% and 15\% increases, respectively. (Note the
non-perturbative piece contributes $\sim 2\times 10^{-7}$ with these
parameter values.)

Another interesting case where the five-form terms could have
contributed is in the calculation of $\eta/s$ in the background dual
to the boost-invariant plasma \cite{Janik:2005zt}. In this
background the tensor $\mathcal T$ is no longer vanishing. Indeed,
$P F_5^2$ is still zero, but now $\nabla F_5$ is of order
$\alpha(u,t)$, the warp factor on the five-sphere. Thus one expects
that the terms $C^2 \nabla F_5^2$ and $(\nabla F_5)^4$ will give a
non-zero contribution (no odd powers appear in the real part of
$\mathcal W$). However, in \cite{Benincasa:2007tp} it was shown that
the field $\alpha(u,t)$ is zero up to an irrelevant term for the
$\eta/s$ computation. The equation of motion for $\alpha$ becomes
(schematically)
$$
\nabla^2 \alpha(u,t)=\gamma(\mathcal O(1)+\mathcal O(\alpha(u,t))+...),
$$
and the only relevant piece of the right is the $\mathcal O(1)$
terms coming from $C^4$. This means that the new terms do not
contribute to the sourcing of $\alpha(u,t)$, and further they will
also not change Einstein's equations since $\alpha(u,t)$ shows up
quadratically there and the above equation determines it to be
$\mathcal O(\alpha'^3)$. We conclude that the five-form terms
contained in $\mathcal W$ cannot change the $\eta/s$ computation in
this framework. Hence the agreement between this computation and the
equilibrium ones is preserved.

Finally, note that the present discussion assumes a simple product
geometry $A_5\times \cM$ and so only applies for thermodynamics in
the absence of a chemical potential. The latter may be included by
considering $R$-charge black holes. However, in this case, the
equations of motion are known to be corrected by the $F_5$ terms
\cite{Paulos:2008tn}. In fact, while the analysis using the only
$C^4$ term seems prohibitively difficult in this case, it was found
that including the five-form produces a particularly simple result.
Hence it will be very interesting to compute $\eta/s$ in this
context.

\acknowledgments{We thank Paolo Benincasa, Alex Buchel, Michael
Green, Krishna Rajagopal and Samuel Vazquez for useful discussions.
Research at Perimeter Institute is supported by the Government of
Canada through Industry Canada and by the Province of Ontario
through the Ministry of Research \& Innovation. RCM also
acknowledges support from an NSERC Discovery grant and funding from
the Canadian Institute for Advanced Research. MP is supported by the
Portuguese Fundacao para a Ciencia e Tecnologia, grant
SFRH/BD/23438/2005. MP also gratefully acknowledges Perimeter
Institute for its hospitality at the beginning of this project.


\begin{thebibliography}{99}
\bibitem{comm1} Note that all the other components vanish on using the lowest order equations of motion.
\end{thebibliography}
\end{document}